# Processing of Crowdsourced Observations of Aircraft in a High Performance Computing Environment

Andrew Weinert
Lincoln Labaoratory
Surveillance Systems
MIT Lincoln Laboratory
andre.weinert@ll.mit.edu

Ngaire Underhill
Lincoln Labaoratory
Surveillance Systems
MIT Lincoln Laboratory
ngaire.underhill@ll.mit.edu

Bilal Gill
Lincoln Labaoratory
Surveillance Systems
MIT Lincoln Laboratory
bilal.gill@ll.mit.edu

Ashley Wicks
Lincoln Labaoratory
Surveillance Systems
MIT Lincoln Laboratory
0000-0003-0937-807X

*Abstract*—As unmanned aircraft systems (UASs) continue to integrate into the U.S. National Airspace System (NAS), there is a need to quantify the risk of airborne collisions between unmanned and manned aircraft to support regulation and standards development. Both regulators and standards developing organizations have made extensive use of Monte Carlo collision risk analysis simulations using probabilistic models of aircraft flight. We've previously determined that the observations of manned aircraft by the OpenSky Network, a community network of ground-based sensors, are appropriate to develop models of the low altitude environment. This works overviews the high performance computing workflow designed and deployed on the Lincoln Laboratory Supercomputing Center to process 3.9 billion observations of aircraft. We then trained the aircraft models using more than 250,000 flight hours at 5,000 feet above ground level or below. A key feature of the workflow is that all the aircraft observations and supporting datasets are available as open source technologies or been released to the public domain.

## I. Introduction

The continuing integration of unmanned aircraft system (UAS) operations into the National Airspace System (NAS) requires new or updated regulations, policies, and technologies to maintain safe and efficient use of the airspace. To help achieve this, regulatory organizations such as the Federal Aviation Administration (FAA) and the International Civil Aviation Organization (ICAO) mandate the use of collision avoidance systems to minimize the risk of a midair collision (MAC) between most manned aircraft (e.g. 14 CFR § 135.180).

Monte Carlo safety simulations and statistical encounter models of aircraft behavior [1] have enabled the FAA to develop, assess, and certify systems to mitigate the risk of airborne collisions. These simulations and models are based on observed aircraft behavior and have been used to design, evaluate, and validate collision avoidance systems deployed on manned aircraft worldwide [2].

### A. Motivation

For assessing the safety of UAS operations, the Monte Carlo simulations need to determine if the UAS would be a hazard to manned aircraft. Therefore there is an inherent need for models that represent how manned aircraft behave. While various models have been developed for decades, many of these models were not designed to model manned aircraft behavior where UAS are likely to operate [3]. In response, new models designed to characterize the low altitude environment are required. In response, we previously identified and determined that the OpenSky Network [4], a community network of ground-based sensors that observe aircraft equipped with Automatic Dependent Surveillance-Broadcast (ADS-B) out, would provide sufficient and appropriate data to develop new models [5]. ADS-B was initially developed and standardized to enable aircraft to leverage satellite signals for precise tracking and navigation. [6, 7]. However, the previous work did not train any models.

### B. Scope

This work considered only how aircraft, observed by the OpenSky Network, within the United States and flying between 50 and 5,000 feet above ground level (AGL) or less. Thus this work does not consider all aircraft, as not all aircraft are equipped with ADS-B. The scope of this work was informed by the needs of FAA UAS Integration Office, along with the activities of the standards development organizations of ASTM F38, RTCA SC-147, and RTCA SC-228. Initial scoping discussions were also informed by the UAS ExCom Science and Research Panel (SARP), an organization chartered under the ExCom Senior Steering Group; however the SARP did not provide a final review of the research.

### C. Objectives and Contributions

We focused on two objectives identified by the aviation community to support integration of UAS into the NAS. First to train a generative statistical model of how manned aircraft behavior at low altitudes. And second to estimate the relative frequency that a UAS would encounter a specific type of aircraft. These contributions are intended to support current and expected UAS safety system development and evaluation and facilitate stakeholder engagement to refine our contributions for policy-related activities.

The primary contribution of this paper is the design and evaluation of the high performance computing (HPC) workflow to train models and complete analyses that support the community's objectives. Refer to previous work [5, 8] for discussion on the assessment of the training data source or how

This material is based upon work supported by the Federal Aviation Administration under Air Force Contract No. FA8702-15-D-0001. Any opinions, findings, conclusions or recommendations expressed in this material are those of the author(s) and do not necessarily reflect the views of the Federal Aviation Administration. Delivered to the U.S. Government with Unlimited Rights, as defined in DFARS Part 252.227-7013 or 7014 (Feb 2014). Notwithstanding any copyright notice, U.S. Government rights in this work are defined by DFARS 252.227-7013 or DFARS 252.227-7014 as detailed above. Use of this work other than as specifically authorized by the U.S. Government may violate any copyrights that exist in this work.

to use the results from this workflow. This paper focus primarily on the use of the Lincoln Laboratory Supercomputing Center (LLSC) [9] to process billions of aircraft observations in a scalable and efficient manner.

## II. STORAGE AND COMPUTE ARCHITECTURE

We first briefly overview the storage and compute infrastructure of the LLSC. The LLSC and its predecessors have been widely used to process aircraft tracks and support aviation research for more than a decade.

### A. Storage and Filesystem

The LLSC High-Performance Computing (HPC) systems have two forms of storage: distributed and central. Distributed storage is comprised of the local storage on each of the compute nodes and this storage is typically used for running database applications. Central storage is implemented using the open-source Lustre parallel file system on a commercial storage array. Lustre provides high performance data access to all the compute nodes, while maintaining the appearance of a single filesystem to the user. The Lustre filesystem is used in most of the largest supercomputers in the world. Specifically, the block size of Lustre is 1MB, thus any file created on the LLSC will take at least 1MB of space.

### B. Compute Infrastructure

The processing described in this paper was conducted on the LLSC HPC system [9]. The system consists of a variety of hardware platforms, but we specifically developed, executed, and evaluated our software using compute nodes based on dual socket Haswell (Intel Xeon E5- 2683 V3 @ 2.0 GHz) processors. Each Haswell processor has 14 cores and can run two threads per core with the Intel Hyper-Threading technology. The Haswell node has 256 GB of memory.

## III. PROCESSING AND RESULTS

This section describes the high performance computing workflow and the results for each step.

### A. Raw Data

A shell script was used to download the raw data archives for a given Monday from the OpenSky Network. Data was organized by day and hour. Both the OpenSky Network and our architecture will create a dedicated directory for a given day, such as 2020-06-22. After extracting the raw data archives, up to 24 comma separated value (csv) files will populate the directory; each hour in UTC time corresponds to a specific file. However, there are a few cases where not every hour of the day was available. The files contain all the abstracted observations of all aircraft for that given hour. For a specific aircraft, observations are updated at least every ten seconds. For this paper, we downloaded 85 Mondays spanning February 2018 to June 2020, totaling 2002 hours.

The size of each hourly file was dependent upon the number of active sensors that hour, the time of day, the quantity of aircraft operations, and the diversity of the operations. Across a given day, the hourly files can range in size by hundreds of megabytes with the maximum file size between 400 and 600 megabytes. Together all the hourly files for a given day currently require about 5-9 gigabytes of storage. We observed that on average the daily storage requirement for 2019 was greater than for 2018.

### B. Organization

Parsing, organizing, and aggregating the raw data for a specific aircraft required high performance computing resources, especially when organizing the data at scale. Many aviation use cases require organizing data and building a track corpus for each specific aircraft. Yet it was unknown how many unique aircraft were observed in a given hour and if a given hourly file has any observations for a specific aircraft. To efficiently organize the raw data, we need to address these unknowns.

We identified unique aircraft by parsing and aggregating the national aircraft registries of the United States, Canada, the Netherlands, and Ireland. Registries were processed for each individual year for 2018-2020. All registries specified the registered aircraft's type (e.g. rotorcraft, fixed wing single-engine, etc.), the registration expiration date, and a global unique hex identifier of the transponder equipped on the aircraft. This identifier is known as the ICAO 24-bit address [10], with ($2^{24}$-2) unique addresses available worldwide. Some of the registries also specified the maximum number of seats for each aircraft.

Using the registries, we created a four tier directory structure to organize the data. The highest level directory corresponds to the year, such as 2019. The next level was organized by twelve general aircraft type, such as fixed wing single-engine, glider, or rotorcraft. The third directory level was based on the number of seats, with each directory representing a range of seats. A dedicated directory was created for aircraft with an unknown number of seats. The lowest level directory was based on the sorted unique ICAO 24-bit addresses. For each seat-based directory, up to 1000 ICAO 24-bit address directories are created. Additionally to address that the four aircraft registries do not contain all registered aircraft globally, a second level directory titled "Unknown" was created and populated with directories corresponding to each hour of data. The top and bottom level directories remained the same as the known aircraft types. The bottom directories for unknown aircraft are generated at runtime.

This hierarchy ensures that there are no more than 1000 directories per level, as recommended by the LLSC, while organizing the data to easily enable comparative analysis between years or different types of aircraft. The hierarchy was also sufficiently deep and wide to support efficient parallel process I/O operations across the entire structure. For example, a full directory path for the first three tiers of the directory hierarchy could be: "2020/Rotorcraft/Seats_001_010/." The directory would contain all the known unique ICAO 24-bit addresses for rotorcraft with 1-10 seats in 2018. Within this directory would be up to 1000 directories, such as "A00C12_A00D20" or "A000D20_A00ECF" This lowest level directory would be used to store all the organized raw data for aircraft with an ICAO 24-bit address. The first hex value was inclusive, but the second hex value was not inclusive.

With a directory structure established, each hourly file was then loaded into memory, parsed, and lightly processed.

Observations with incomplete or missing position reports were removed, along with any observations outside a user-defined geographic polygon. The default polygon, illustrated by Figure 1, was a convex hull with a buffer of 60 nautical mile around approximately North America, Central America, the Caribbean, and Hawaii. Units were also converted to U.S. aviation units. The country polygons were sourced from Natural Earth, a public domain map dataset [11].

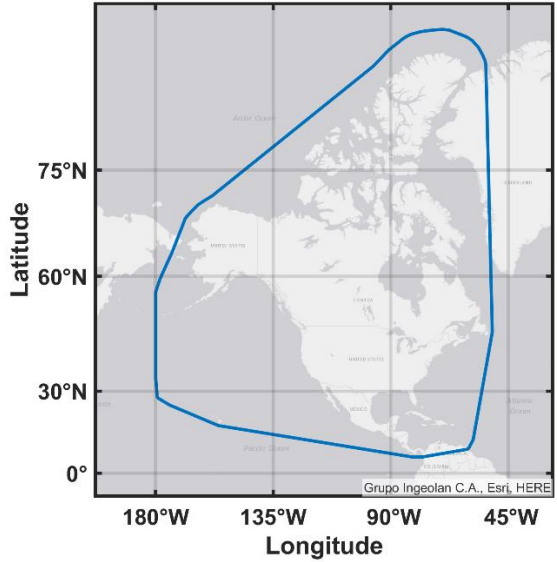

Fig. 1. Geospatial filtering polygon used for coarse organization.

After processing, an individual csv file was created for each hour for each aircraft. For example suppose an aircraft with an address of A00C12 was observed from 05:00 to 06:30 UTC on 2020-03-16. The following two files would be created:

- /2020/Rotorcraft/Seats_001_010/A00C12_A00D20/2020-03-16_05_A00CDE.csv
- /2020/Rotorcraft/Seats_001_010/A00C12_A00D20/2020-03-16_06_A00CDE.csv

Specifically for the 85 Mondays across the three years, 2214 directories were generated across the first three tiers of the hierarchy and 802,159 directories were created in total across the entire hierarchy. Of these, 770,661 directories were non-empty. The majority of the directories were created within the unknown aircraft type directories. As overviewed by Tables 1 and 2, about 3.9 billion raw observations were organized, with about 1.4 billion observations available after filtering. There was a 15% annual percent increase in observations per hour from 2018 to 2019. However, a 50% percent decrease in the average number of observations per hour was observed when comparing 2020 to 2019; this could be attributed to the COVID-19 pandemic. This worldwide incident sharply curtailed travel, especially travel between countries.

This reduction in travel was reflected in the amount of data filtered using the geospatial polygon. In 2018 and 2019, about 41-44% of observations were filtered based on their location. However, only 27% of observations were filtered for observations from March to June 2020. Conversely, the amount of observations removed due to quality control did not significantly vary in 2020, as 26%, 20%, and 25% were removed for 2018, 2019, and 2020.

TABLE I. ORGANIZING STATISTICS - TOTAL

| Year | Hours | Raw | Organized |
|---|---|---|---|
| 2018 | 724 | 1,539,058,315 | 502, 407, 955 |
| 2019 | 942 | 2,131,412,447 | 761, 908, 350 |
| 2020 | 336 | 320,233,440 | 155, 814, 012 |
| Total | 2002 | 3,990,704,202 | 1,420,130,317 |

TABLE II. ORGANIZING STATISTICS – HOURLY AVERAGE

| Year | Hours | Raw | Organized |
|---|---|---|---|
| 2018 | 724 | 2,125,771 | 693,934 |
| 2019 | 942 | 2,262,646 | 808,820 |
| 2020 | 336 | 953,076 | 463,732 |

These results were generated using 512 CPUs across 2002 tasks, where each task corresponded to a specific hourly file. Tasks were uniformly distributed across CPUs, a dynamic self-scheduling parallelization approach was not implemented. Each task required on average 626 seconds to execute, with a median time of 538 seconds. The maximum and minimum times to complete a task were 2153 and 23 seconds. Across all tasks, about 348 hours of total compute time was required to parse and filter the 85 days of data. It is expected that if the geospatial filtering was relaxed and observations from Europe were not removed, that the compute time would increase due to increase demands on creating and writing to hourly files for each aircraft.

*C. Archive Organized Data*

Since files were created for every hour for each unique aircraft, tens of millions of small files less than 1 megabyte in size were created. This was problematic as small files typically use a single object storage target, thus serializing access to the data. Additionally, in a cluster environment, hundreds or thousands of concurrent, parallel processes accessing small files can lead to significantly large random I/O patterns for file access and generates massive amounts of networks traffic. This results in increased latency for file access, higher network traffic and significantly slows down I/O and consequently causes degradation in overall application performance. While this approach to data organization may provide acceptable performance on a laptop or desktop computer, it was unsuitable for use in a shared, distributed HPC system.

In response, we created zip archives for each of the bottom directories. In a new parent directory, we replicated the first three tiers of the directory hierarchy from the previous step. Then instead of creating directories based on the ICAO 24-bit addresses, we archiving each directory with the hourly csv files from the previous organization step. We then removed the hourly csv files from storage. This was achieved using

LLMapReduce [12], with a task created for each of the 770,661 non-empty bottom level directories. Similar to the previous organization step, all tasks were completed in a few hours but with no optimization for load balancing. The performance of this step could be improved by distributing tasks based on the number of files in the directories or the estimated size the output archive.

A key advantage to archiving the organized data, is that the archives can be updated with new data as it becomes available. If the geospatial filtering parameters and aircraft registry data doesn't change, only new Open Sky data needs to be organized. Once organized into individual csv files, LLMapReduce can be used again to update the existing archives. This substantially reduces the computational and storage requirements to process new data.

*D. Process and Interpolate Data*

The archived data can now be segmented, have outliers removed, and interpolated. Additionally above ground level altitude was calculated, airspace class was identified, and dynamic rates (e.g. vertical rate) were calculated. We also split the raw data into track segments based on unique position updates and time between updates. This ensures that each segment does not include significantly interpolated or extrapolated observations. Track segments without ten points are removed. Figure 2 illustrates the track segments for a FAA registered fixed wing multi-engine aircraft from March to June 2020. Note that segment length can vary from tens to hundreds of nautical miles long. Track segment length was dependent upon the aircraft type, availability of active OpenSky Network sensors, and nearby terrain. However, the ability to generate track segments that span multiple states represents a substantial improvement over previous processing approaches for development of aircraft behavior models.

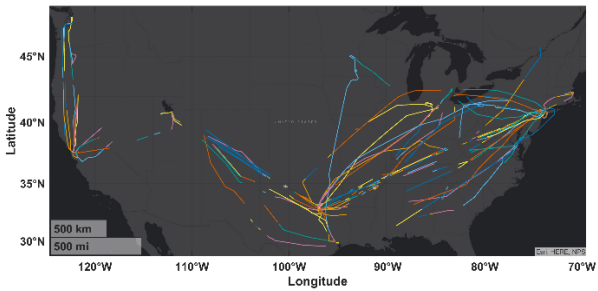

Fig. 2. Track segements for a FAA registered fixed wing multi-engine aircraft from March to June 2020.

Then for each segment we detect altitude outliers using a 1.5 scaled median absolute deviations approach and smooth the track using a Gaussian-weight average filter with a 30-second time window. Dynamic rates, such as acceleration, are calculated using a numerical gradient. Outliers are then detected and removed based on these rates. Outlier thresholds were based on aircraft type. For example, the speeds greater than 250 knots were considered outliers for rotorcraft, but fixed wing multi-engine aircraft had a threshold of 600 knots. The tracks were then interpolated to a regular one second interval.

Lastly, we estimated the above ground level altitude using digital elevation models. This altitude estimation was the most computationally intensive component of the entire workflow. It consists of loading into memory and interpolating SRTM3 or NOAA GLOBE [13] digital elevation models (DEMs) to determine the elevation for each interpolated track segment position. To reduce the computational load prior to processing the terrain data, it was determined using a C++ based polygon test to identify which track segment positions are over the ocean, as defined by Natural Earth Data. Points are over the ocean are assumed to have an elevation of 0 feet mean sea level and their elevation are not estimated using the DEMs.

For the 85 days of organized data, approximately 900,000,000 interpolated track segments were generated. For each aircraft in a given year, a single csv was generated containing all the computed segments. In total across the three years, 619,337 files were generated. As these files contained significantly more rows and columns than when organizing the raw data, the majority of these final files were greater than 1 MB in size. The output of this step did not face any significant storage block size challenges.

Similar to the previous step, tasks were created based on the bottom tier of the directory hierarchy. Specifically for processing, parallel tasks were created for each archive. During processing, archives were extracted to a temporary directory while the final output was stored in standard memory.

## IV. DISCUSSION AND APPLICATIONS

Given the processed data, this section overviews two applications on how to exploit and dissemination the data to inform and support the aviation safety community.

*A. Distribution of Aircraft Models*

As the aircraft type was identified when organizing the raw data, it was a straightforward task to estimate the observed distribution of aircraft types per hour.

TABLE III. AVERAGE AIRCRAFT TYPE DISTRIBUTION PER HOUR

| Year | *Fixed Wing Multi Engine* | *Fixed Wing Single Engine* | *Rotorcraft* | *Unknown* |
|------|---------------------------|----------------------------|--------------|-----------|
| 2018 | 71% | 8% | 2% | 19% |
| 2019 | 76% | 11% | 3% | 10% |
| 2020 | 62% | 21% | 7% | 10% |

These distributions are not reflective of all aircraft operations in the United States, as not all aircraft are observed by the OpenSky Network. The distributions were also calculated independently for each aircraft type, so the yearly (row) percentages may not sum to 100%. Furthermore the relatively low percentage of unknown aircraft was due to the geospatial filtering when organizing the raw data. If the same aircraft registries were used by the filtering was change to only include tracks in Europe, the percentage of unknown aircraft would likely significantly rise.

This analysis can be extended by identifying specific aircraft manufactures and models, such as Boeing 777. However, the manufacturer and model information are not consistent within an aircraft registry nor across different registries. For example, entries of "Cessna 172," "Textron Cessna 172," and "Textron C172" all refer to the same aircraft model. One possible explanation for the differences between entries is that Cessna used to be an independent aircraft manufacturer and then eventually was acquired by Textron. Depending on the year of registration, the name of the aircraft may differ but the size and performance of the aircraft remains constant.

Since over 300,000 aircraft with unique ICAO 24-bit addresses were identified annually across the aircraft registries, parsing and organizing the aircraft models can be formulated as a traditional natural language processing problem. Parsing the aircraft registries differs from a common problem of parsing aviation incident or safety reports [14, 15, 16] due to the reduced word count of the registries and the structured format of the registries. Future work will focus on using fuzzy string matching to identify similar aircraft.

### B. Statistical Models of Aircraft Behavior

For many aviation safety studies, manned aircraft behavior is represented using MIT Lincoln Laboratory encounter models. Each encounter model is a Bayesian Network, a generative statistical model that mathematically represents aircraft behavior during close or safety critical encounters, such as near midair collisions. The development of the modern models started in 2008 [1], with significant updates in 2013 [17] and 2018 [18]. All the models were trained using the LLSC [9] or its predecessors. The most widely used of these models were trained using observations collected by ground-based secondary surveillance radars from the 84th Radar Evaluation Squadron (RADES) network.

Aircraft observations by the RADES network are based on Mode 3A/C, an identification friend or foe technology that provides less metadata than ADS-B. Notably aircraft type or model cannot be explicitly correlated or identified with specific aircraft tracks. Instead, we filtered the RADES observations based on the flying rules reported by the aircraft. However, this type of filtering is not unique to the RADES data, it is also supported by the OpenSky Network data.

Additionally, due to the performance of the RADES sensors, we filtered out any observations below 500 feet AGL due to position uncertainties associated with radar time of arrival measurements. Observations of ADS-B equipped aircraft by the OpenSky Network differ because ADS-B enables aircraft to broadcast the aircraft's estimate of their own location, which is often based on precise GNSS measurements. The improved position reporting of ADS-B enabled the new OpenSky Network-based models to be trained with an altitude floor of 50 feet AGL, instead of 500. Specifically, three new statistical models of aircraft behavior were trained, each for a different aircraft type of fixed wing multi-engine, fixed wing single-engine, and rotorcraft. A key advantage to these models is the data reduction and dimensionality reduction. A model was created for each of the three aircraft types and stored as a human readable text file. Each file requires approximately just 0.5 megabytes. This a significant reduction from the hundreds of gigabytes used to store the original 85 days of data.

Table IV reports the quantity of data used to train each model. For example, the rotorcraft model was trained from about 25,000 flight hours over 85 days. However, like the RADES-based model, these models do not represent the geospatial nor temporal distribution of the training data. For example, a limitation of these models is that they do not inform if more aircraft were observed in New York City than Los Angeles.

TABLE IV. PROCESSED FLIGH HOURS BETWEEN 50 AND 5,000 FEET AGL

| Year | Fixed Wing Multi Engine | Fixed Wing Single Engine | Rotorcraft | Total |
|---|---|---|---|---|
| 2018 | 31,669 | 34,043 | 6,133 | 71,845 |
| 2019 | 56,176 | 63,826 | 12,837 | 132,839 |
| 2020 | 12,032 | 30,387 | 6,984 | 49,403 |
| Total | 99,877 | 128,255 | 25,954 | 254,097 |

Figures 3 and 4 illustrate the altitude and speed distributions of the various models. There are multiple RADES-based models, and these figures only illustrate the model trained on aircraft operating under visual flight rules (VFR) [17]. These figures illustrate how different aircraft behave, such as rotorcraft flying relatively lower and slower than fixed wing multi-engine aircraft. Also note that the RADES-based model has no altitude observations below 500 feet AGL, whereas 18% of the approximately 25,000 rotorcraft flight hours were observed at 50-500 feet AGL.

It has not been assessed if the OpenSky Network-based models can be used a surrogates for other aircraft types or operations. Additionally the new models do not fully supersede the existing RADES-based models, as each models represent different varieties of aircraft behavior.

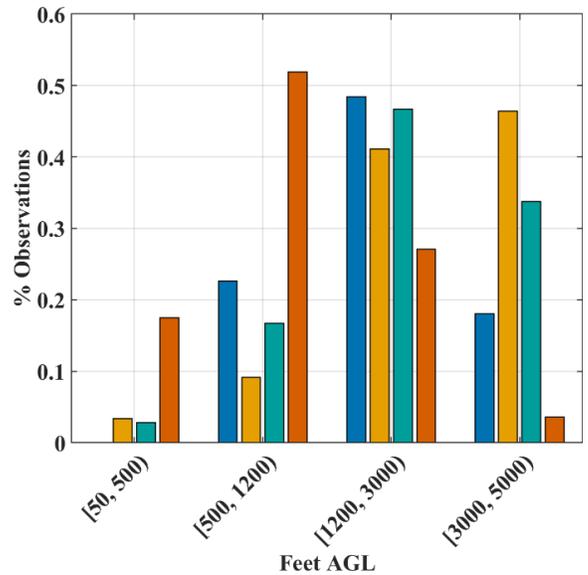

Fig. 3. Altitude distribution of statistical models. LEGEND: Blue: RADES-Based VFR; Orange: OpenSky Network-based Fixed Wing Multi-Engine;

Blueish Green: OpenSky Network-based Fixed Wing Single-Engine; Vermillion: OpenSky Network-based Rotorcraft

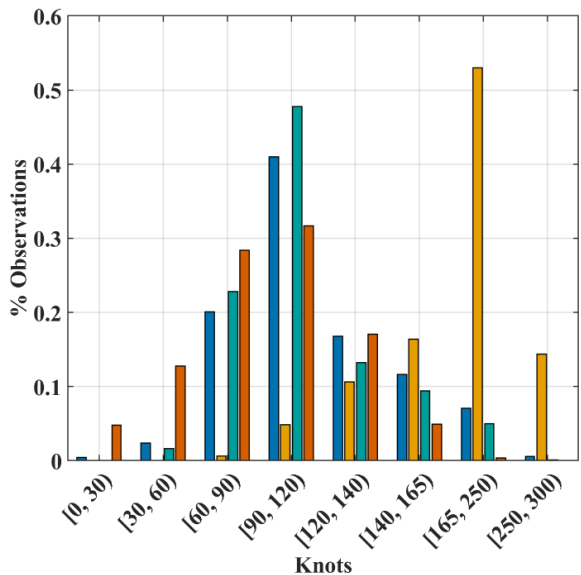

Fig. 4. Speed distribution of statistical models. LEGEND: Blue: RADES-Based VFR; Orange: OpenSky Network-based Fixed Wing Multi-Engine; Blueish Green: OpenSky Network-based Fixed Wing Single-Engine; Vermillion: OpenSky Network-based Rotorcraft

## V. CONCLUSION

We developed and deployed a workflow to efficiently organize, archive, and process aircraft tracks observed by the OpenSky Network. Many of the capabilities described in this paper have been, or are in the process of being, transitioned as open source software under permissive open source licenses. On GitHub.com, please refer to the MIT Lincoln Laboratory (@mit-ll) and Airspace Encounter Models (@Airspace-Encounter-Models) organizations.

ACKNOWLEDGMENT

We greatly appreciate the support and assistance provided by Sabrina Saunders-Hodge, Richard Lin, and Adam Hendrickson from the Federal Aviation Administration. We also would like to thank fellow colleagues Dr. Rodney Cole, Matt Edwards, and Wes Olson.

REFERENCES

[1] M. Kochenderfer, M. Edwards, L. Espindle, J. Kuchar and J. D. Griffith, "Airspace encounter models for estimating collision," *Journal of Guidance, Control, and Dynamics,* vol. 33, no. 2, pp. 487-499, 2010.

[2] L. P. Espindle, J. D. Griffith and J. K. Kuchar, "Safety Analysis of Upgrading to TCAS Version 7.1 Using the 2008 U.S. Correlated Encounter Model," MIT Lincoln Laboratory, Lexington, MA, USA, 2009.

[3] A. Weinert, S. Campbell, A. Vela, D. Schuldt and J. Kurucar, "Well-Clear Recommendation for Small Unmanned Aircraft Systems Based on Unmitigated Collision Risk," *AIAA Journal of Air Transportation,* vol. 26, no. 3, pp. 113-122, 2018.

[4] M. Schäfer, M. Strohmeier, V. Lenders, I. Martinovic and M. Wilhelm, "Bringing Up OpenSky: A Large-scale ADS-B Sensor Network for Research," in *13th IEEE/ACM International Symposium on Information Processing in Sensor Networks (IPSN)*, Berlin, 2014.

[5] A. Weinert, N. Underhill and A. Wicks, "Developing a Low Altitude Manned Encounter Model Using ADS-B Observations," 2019 IEEE Aerospace Conference, Big Sky, MT, 2019.

[6] G. Donohue, "Vision on aviation surveillance systems," in *Proceedings International Radar Conference*, Alexandria, VA, USA, 1995.

[7] V. A. Orlando, "ADS-Mode S: Initial System Description," Massachusetts Institute of Technology Lincoln Laboratory, Lexington, MA, USA, 1993.

[8] A. Weinert, M. Edwards and S. M. Katz, "Representative Small UAS Trajectories for Encounter Modeling," in *AIAA Scitech 2020 Forum*, Orlando, FL, USA, 2020.

[9] A. Reuther, J. Kepner, C. Byun, S. Samsi, W. Arcand, D. Bestor, B. Bergeron, V. Gadepally, H. Michael, M. Hubbell, M. Jones, A. Klein, L. Milechin, J. Mullen, A. Prout, A. Rosa, C. Yee and P. Michaleas, "Interactive Supercomputing on 40,000 Cores for Machine Learning and Data Analysis," in *2018 IEEE High Performance extreme Computing Conference (HPEC)*, Waltham, MA, USA, 2018.

[10] R. M. Trim, "Mode S: an introduction and overview (secondary surveillance radar)," *Electronics & Communication Engineering Journal,* vol. 2, no. 2, pp. 53-59, April 1990.

[11] N. V. Kelso and T. Patterson, "Introducing natural earth data-naturalearthdata.com," *Geographia Technica,* vol. 5, pp. 82-89, 2010.

[12] C. Byun, J. Kepner, W. Arcand, D. Bestor, B. Bergeron, V. Gadepally, M. Hubbell, P. Michaleas, J. Mullen, A. Prout, A. Rosa, C. Yee and A. Reuther, "LLMapReduce: Multi-level map-reduce for high performance data analysis," in *2016 IEEE High Performance Extreme Computing Conference (HPEC)*, Waltham, MA, USA, 2016.

[13] D. A. Hastings, P. K. Dunbar, G. M. Elphingstone, M. Bootz, H. Murakami, H. Maruyama, H. Masaharu, P. Holland, J. Payne, N. A. Bryant, T. L. Logan, J. -P. Muller, G. Schreier and J. S. MacDonald, "The Global Land One-kilometer Base Elevation (GLOBE) Digital Elevation Model, Version 1.0," National Oceanic and Atmospheric Administration, National Geophysical Data Center, Boulder, Colorado, 1999.

[14] K. D. Kuhn, "Using structural topic modeling to identify latent topics and trends in aviation incident reports," *Transportation Research Part C: Emerging Technologies,* vol. 87, pp. 105-122, February 2018.

[15] S. D. Robinson, "Temporal topic modeling applied to aviation safety reports: A subject matter expert review," *Safety Science,* vol. 116, pp. 275-286, July 2019.

[16] R. M. Keller, "Ontologies for aviation data management," in *2016 IEEE/AIAA 35th Digital Avionics Systems Conference (DASC)*, Sacramento, CA, USA, 2016.

[17] A. J. Weinert, E. P. Harkleroad, J. D. Griffith, M. W. Edwards and M. J. Kochenderfer, "Uncorrelated Encounter Model of the National Airspace System, Version 2.0," Massachusetts Institute of Technology, Lincoln Laboratory, Lexington, 2013.

[18] N. Underhill, E. Harkleroad, R. Gundel, A. Weinert, E. Maki and M. Edwards, "Correlated Encounter Model for Cooperative Aircraft in the National Airspace System Version 2.0," Massachusetts Institute of Technology, Lincoln Laboratory, Lexington, 2018.